\documentclass[12pt,a4paper]{article}

\usepackage{epsfig}
\usepackage{amsmath}
\begin{document}

\begin{center}
\large
{\bf Raman Scattering Characterization and Electron Phonon 
Coupling Strength for MeV implanted InP(111)}
\end{center}
\normalsize
\vspace*{0.1cm}
 \centerline {D. Paramanik and S. Varma$^{*}$}
\vspace*{0.1cm}
\small
\centerline{\it Institute of Physics, Bhubaneswar - 751005, India.}
\vspace*{0.3cm}
\begin{center}
{\bf Abstract}
\end{center}

\vspace*{-.1cm}
  Structural modifications in InP(111) due to 1.5~MeV implantation of 
Sb have been characterized using first order and second order Raman 
spectroscopy. With both Longitudinal Optical (LO) and Transverse Optical 
(TO) modes allowed for InP(111), we have investigated the evolution of 
both these modes as a function of fluence. Intensity, linewidth and 
shifts of the phonons, for both first order and second order Raman 
modes, display the increase in damage in the lattice  with increasing
fluence. The results suggest that the presence of a charge layer in the 
vicinity of the surface may be effecting the first order Raman data.
A LO phonon-plasmon coupled mode, due to the charge layer, 
has also been observed that becomes sharper and more intense
with increasing fluence.  Results also show the presence of tensile
stress along with the coexistence of crystalline InP regions and 
amorphous zones in the lattice. Consequently phonon confinement is 
observed.  Phonon Confinement model (PCM) has been applied here to 
estimate the coherence length and the size of nano-crystalline zones 
in InP lattice after implantation. A crystalline/ amorphous (c/a) 
phase transition is observed at the fluence of $1\times10^{14} ions/cm^{2}$. 
The electron-phonon coupling strength has been measured by utilizing the 
second order Raman modes. This coupling strength is seen to decrease 
as the nano-crystalline zones, in the implanted lattice, become smaller.

\vspace*{1.0cm}


\vspace*{0.1cm}
\hrule width 220pt depth 1pt height  0.3pt
\vspace*{0.1cm}
\noindent $^*${\em Corresponding author: shikha@iopb.res.in, tel:91-674-2301058,
FAX: 91-674-2300142}        

\newpage
\normalsize
\section{Introduction:}

 Unique properties of InP have attracted enormous interest. It is being widely
applied in high speed electronic and optoelectronic devices due to its attractive
electronic properties \cite{str,hum,lam} as well as its excellent lattice match 
with low band gap alloys like GaInAs, GaInAsP, AlGaInAs etc. GaInAsP/InP based 
photo-diodes operate in low loss window of silica fibers with high quantum 
efficiency and fast response time. Due to its excellent physical properties like
high thermal conductivity, high peak velocities for electrons and holes, InP 
is considered an important semiconductor material and it is being prominently 
utilized in the devices for high electron mobility transistors, high efficiency 
and high speed quantum well lasers, photo-detectors, photonic integrated circuits 
etc. InP is also preferred, over GaAs based devices, for Millimeter-wave sources 
and amplifiers due to its low noise and higher efficiency operations in high frequency 
regime. Junction field effect transistors fabricated on InP display high 
performance.  Sb is considered an important dopant because of its role in the 
development of field effect transistors and infrared detectors \cite{temp}.  
Sb has also found applications as isoelectronic dopant for trapping charge 
carriers in InP \cite{bishop}.
Ion-beam implantation techniques are extensively utilized for device processing in 
semiconductor industry. This is an attractive technique as it provides  well 
controlled and sharp dopant profiles. Due to its low thermal stability, MeV 
ion implantation is a prominent way to introduce and dope the materials in InP. 
MeV ion implantation offers a means of doping relatively thicker buried layers 
with modified properties as well as modifications of vertically limited layers 
and quantum well structures. The increased density in VLSI circuits also makes
the technological applications of the ion implantation, especially in MeV energy 
range, increasingly important. MeV implantation, however, can also produce damage 
and can cause severe modifications in the material depending on the nature and 
the energy of the impinging ion, and the implantation dose \cite{tam1,sdey1}.
Hence, for ion implantation to be a viable candidate for the development in 
semiconductor technology, it is important to understand and characterize the
bulk and surface disorder. In earlier studies we have applied Scanning Probe
microscopy to investigate the surface modifications on Si(100) \cite{sdey3} 
and InP(111) \cite{par}. For investigating bulk lattice modifications Raman 
scattering intensity, half width and peak shift of the zone center phonons
are considered very sensitive. Our earlier study of MeV implantation in Si 
\cite{sdey2} had shown that the Raman scattering is a powerful technique for 
investigating and monitoring the implantation induced lattice modifications.

During implantation, a projectile while moving forward produces vacancies and 
interstitials, loses energy primarily due to electronic encounters and is finally 
deposited at its range governed by its mass and implant energy \cite{chu}. 
At MeV energies, nuclear energy loss (S$_n$) processes are expected to be 
dominantly responsible for the material modifications. Defects and strains 
can get produced, via $S_n$, causing the modifications in properties of host 
material \cite{gkm,sdey1}. Formation of defects as well as the presence of impurity 
atoms can lead to stress in the planes of the single crystal or changes in the force 
constants. Corresponding shifts in the phonon frequencies are reflected in the Raman 
spectra. Moreover, the damage produced due to defects can cause phonon confinement
leading to a reduced phonon coherence length as k=0 selection rule is relaxed, giving 
rise to an asymmetric broadening in Raman peak. The Raman spectrum thus contains 
signature of both: the stress and the reduced phonon coherence length due to disorder 
in the lattice. Raman Scattering, thus, is a powerful tool for investigating and 
monitoring the radiation induced lattice modifications during ion implantation.

For III-V semiconductors like InP and other zinc- blend structures, the first 
order Raman spectrum consists of usually two Raman modes corresponding to the 
Longitudinal Optical (LO) and Transverse Optical (TO) phonons associated with 
the Brillouin zone center. The zone center phonons corresponding to TO and LO 
modes obey the following selection rules for zinc blend type crystals. For the 
scattering by (111) face, both LO and TO modes are allowed in Raman scattering. 
However, only TO mode is allowed for scattering by (110) face whereas only LO
mode is allowed for (100) face. Both LO and TO modes being 
allowed for InP(111), we have been able to investigate the evolution of both 
these modes as a function of ion fluence after MeV implantation. A few studies
have investigated the modifications in InP crystals after implantations by Raman 
scattering technique\cite{artus,artus1,tiginyanu,santhakumar,ibanez,cusco,bedel,
soon,cusco1, ramarao}. However, all these studies investigated the (100) InP 
where TO mode is forbidden. Moreover, in all these studies keV implantation 
energies were applied for lattice modifications. There is one report on Raman 
scattering study of ion modification in InP(111) but for low energy keV 
implantation with He ions \cite{tiginyanu}. 

For InP, phonon dispersion studies
indicate \cite{borcherds,borcherds2} that LO branch is almost as flat as TO 
branch for the entire Brillouin zone. Moreover, Overlap Valence Shell Model 
\cite{borcherds2} 
results display two very sharp and well separated features corresponding to 
second order 2LO and 2TO phonon modes with high density of states around 
TO($\Gamma$) and LO($\Gamma$). Consequently, second order phonon studies 
of InP can be effectively undertaken. These second order modes are 
considered to be more sensitive probes, compared to first order, of 
lattice disorder, stress and lattice modifications after implantation 
\cite{artus1}. However, very few studies have investigated these second 
order Raman modes \cite{artus1,ibanez,cusco,bedel,cusco1}. Moreover, 
all of these studies have investigated InP(100) after keV implantations. 
There are no studies in literature where second order Raman modes have been 
investigated after keV or MeV implantation of InP(111) crystals. 

At the surface of III-V semiconductor compounds a space charge layer, 
concentration of high mobility free-electron gas, may exist. 
If the frequency of this free-electron plasma excitations
- plasmons is close to the frequency of LO phonons, the two excitations can
interact via their macroscopic electric fields and can form LO phonon-plasmon
coupled (LOPC) mode. In n-type InP(100), an intense and well defined $L^-$ 
feature due to LOPC mode has been observed \cite{artus}. A few studies have 
investigated the scattering of light by free carriers in zinc- blende semiconductors 
\cite{abs} and some models have been proposed that analyze the coupling between 
LO phonons and plasmons. However, there are very few reports for doped InP 
crystals \cite{artus} and none of these studies focus on the evolution of LO-plasmon
coupling in InP(111) after implantation.  The presence of charge layer also
results in a smaller scattering volume near the surface and effects the 
characteristics of first order LO Raman mode. The second order Raman modes do 
not suffer from this drawback.

 Electron-phonon coupling is a very important factor in understanding the 
non-linear optical properties of the crystalline materials. For polar 
materials, the dominant coupling is  Fr\"{o}hlich type between the field 
induced by the vibrational modes and the electronic charge density \cite{ali}. 
Systematic trends in the electron-phonon coupling strengths, with size, 
have been seen for Quantum Dots of InP by utilizing second order Raman Scattering
\cite{seong}. However, there are no studies in literature where electron-phonon 
coupling strengths have been measured after ion implantation.

In the present study we have utilized first and second order Raman scattering 
phonon spectra to investigate the lattice modifications in InP(111) after 
implantation with 1.5 MeV $Sb^{2+}$ ions. We have studied the evolution
of both TO and LO modes as a function of ion fluence. Even at the fluence of
1$\times10^{11} ~ions/cm^2$, modifications in first order as well as second
order modes demonstrate the initiation of some damage in InP(111) lattice.
Once this occurs, a coexistence of nanocrystalline regions and amorphous 
zones is observed which leads to the confinement of phonons. An estimation 
of the phonon coherence length has been done by applying the Phonon Confinement 
Model (PCM) \cite{ley} to the first order Raman peak. Existence of a completely 
amorphous lattice is noticed for a fluence  of 1$\times10^{14}ions/cm^2$ and
higher.

The evolution of LOPC mode, L$^{-1}$, has also been studied as a function of
ion fluence.  We have also investigated here the effect of implantation on the 
electron phonon coupling in InP(111) by utilizing the second order Raman modes. 
We find that the strength of coupling reduces as the size of nano-crystalline
zones, in the implanted lattice, decrease.

Experimental procedures and results are discussed in section~2 and ~3, 
respectively.  Conclusions are presented in section~4.

\section{Experimental}

A mirror polished (111)-oriented InP single crystal wafer
 was used in the present study. The samples were implanted at room 
temperature with a scanned beam of 1.5 MeV Sb$^{2+}$ ions at various 
fluences ranging from  1$\times 10^{11}$ to 5$\times10^{15} ~ions/cm^2$.
The average Sb flux was 0.002 to 0.02 $\mu$A/cm$^2$. This current was
measured directly on the target after suppressing the secondary
electrons by applying a negative bias of 200V to a suppressor
assembly around the target. The implantation were performed with
the samples oriented 7$^o$ off-normal to the incident beam to
avoid channeling effects. Monte Carlo simulations were performed 
for 1.5~MeV Sb implantation in InP using the SRIM'03 code and
the mean projected range of Sb-ion distribution
was found to be 400~nm \cite{srim}. 

Raman scattering measurements were performed using a SPEX 1877E
Triplemate Spectrometer with a liquid nitrogen cooled, charged coupled
device array. The laser power was controlled to avoid laser
annealing effect on the sample. Raman experiments were carried out
at room temperature using the 514~nm line of an argon ion laser in
the backscattering geometry. At this wavelength the penetration depth
of the light is estimated to be about 100~nm. All the spectra were
acquired in the backscattering geometry.

\section {Results and Discussion}

 Figure~1 shows the as-implanted first order Raman Spectra from the 
InP(111) samples implanted at various fluences. The spectrum from a 
virgin (un-implanted) InP is also shown.  The spectra have
been shifted vertically for clarity, but the intensity scale is the
same for all the spectra. The spectrum of the virgin InP (Fig.1a) shows
the characteristic LO and
TO Raman modes of crystalline InP(111) \cite{pin}. The features at 
305~cm$^{-1}$ and at 347~cm$^{-1}$ are assigned to the TO and the
LO phonon modes, respectively. For the virgin InP(111) we also 
observe an additional L$^-$ mode, near 320~cm$^{-1}$, as a shoulder 
near TO mode. This broad phonon feature can be assigned as L$^-$ LO -
phonon plasmon coupled (LOPC) mode. It is well known that in polar
semiconductors, like InP, free charge couples with LO modes and
forms LOPC-like modes\cite{abs}. For InP(100) where TO mode is forbidden, 
L$^-$ mode has been observed as sharp feature near 306~cm$^{-1}$ 
\cite{soon} as well as at 308~cm$^{-1}$ \cite{moh}.

The sequence of spectra in Fig.~1 gradually evolve
with increasing Sb fluence.  The Raman spectrum from a sample 
implanted with $1\times10^{11}~ions/cm^2$ shows broader LO and TO 
modes (Fig.~1b) compared to the virgin sample. Both the features 
now also exhibit decreased intensity and increased asymmetricity. 
Moreover, both LO and TO modes, after implantation, are shifted 
towards lower wave numbers compared to the modes in the virgin InP. 
All these changes reflect the modifications in the InP lattice due 
to the defects created during implantation. With increasing fluence,
we observe a further decrease in intensity of the TO and LO modes. 
For n-type InP, formation of a charge layer, near surface 
\cite{artus1,pin}, can result in a smaller scattering volume, 
thus effecting the intensity of LO mode. 

  We observe softening as well as asymmetrical broadening in both 
LO and TO modes with increasing fluence. Both the modes exhibit 
shifts towards lower wave numbers. The shifts 
in the peak positions of Raman spectra can be affected by the residual 
stress as well as by phonon confinement. The contributions due to 
the two effects can however be deconvoluted \cite{huang}.
Spatial Correlation model related to {\it k}-vector relaxation 
induced damage shows \cite{tiong} that when disorder is introduced 
into the crystal lattice by implantation, the correlation function
of the phonon-vibrational modes become finite due to the induced 
defects and consequently the momentum $k=0$ selection rule is relaxed. 
Consequently, the phonon modes shift qualitatively to lower frequencies 
and broaden asymmetrically as the ion fluence is increased \cite{sdey2}. 
The reduction in the intensity, shifts towards lower frequencies 
as well as the asymmetrical broadening of the features with increasing
ion fluence, as observed in Fig.~1, are due to the residual defects 
created in InP lattice via implantation. Accordingly, these two features
are also referred to as DALO and DATO, respectively, for disorder activated 
(LO) and (TO) modes. For the fluences of 1$\times10^{14}~ ions/cm^2$ and
higher, no DALO or DATO modes are observed in Fig.~1 suggesting that at 
this stage InP lattice undergoes crystalline/amorphous (c/a) phase 
transition and becomes amorphized. Our results are in contrast to the 
studies of 150~keV Si$^+$ \cite{cusco} or 80~keV Mg \cite{ibanez}
implantation in InP(100) where no noticeable changes compared to the 
virgin were seen upto the fluence of 1$\times10^{12} ions/cm^2$ 
and the first signatures of disorder were observed after the fluence of 
5$\times10^{12} ions/cm^2$. In contrast to these studies, for keV 
implantation of Zn in InP(100), c/a transition was seen at a fluence 
of 1$\times10^{13} ions/cm^2$ \cite{bedel}, which is surprisingly 
lower than the fluence of c/a transition in the present study.
For 2~MeV Se implantation in InP(100), some damage after 
1$\times10^{12} ions/cm^2$ has been reported using channeling 
experiments\cite{wesch}. 
 
 The overall Raman scattering, after implantation, is influenced by the 
presence of the damaged/amorphised zones that get formed in the crystalline
lattice. With increasing fluence, the crystalline regions reduce in size 
whereas the amorphous and damaged zones grow bigger. The disorder is usually 
a consequence of damage zones that form due to the point defects and/or dislocations 
created after MeV implantations. At high fluences 
the damaged zones in the lattice begin to overlap and finally at sufficiently 
high fluences c/a phase transition may occur leading to
a completely amorphised lattice \cite{wemple}.  

Figure~2 displays the quantitative variations, as a function of ion 
fluence, in intensity, linewidth and position of first order Raman 
modes of InP(111) after Sb implantation. In Fig.~2a the Raman intensity 
for TO and LO modes, normalized with respect to the intensity from 
the virgin sample, are shown. For both the modes, we observe a decrease
in intensity for increasing fluence. For TO mode, however, the
decrease is larger and faster. The  widths (FWHM) for both the modes
are shown in Figure~2b. We observe an increase in the linewidths
for both the modes as the fluence is increased. Again, the increase 
is more pronounced in TO mode. The widths of TO and LO modes, after 
the fluence of $1\times10^{13}~ions/cm^2$, are 30 and 20~cm$^{-1}$, 
respectively.  In contrast, much smaller and same linewidths for LO 
and TO modes ($\sim$ 7~cm$^{-1}$) have been observed after keV 
implantation at this fluence \cite{tiginyanu}. 

The positions of the LO and TO modes are shown in Figs.~2c and 2d, 
respectively. Although for the virgin sample the positions for LO 
and TO modes were 347 and 305~cm$^{-1}$, respectively, after the 
fluence of $1\times10^{13}~ions/cm^2$ these mode appear at 339 and 
298~cm$^{-1}$, respectively.  The overall shift ($\omega$) is larger 
and increases more rapidly for LO mode. These shifts 
indicate that the lattice is under stress due to the creation of defects 
after implantation. Moreover, the shift of LO and TO modes towards 
the lower wave numbers suggest that the stress is of tensile nature. 
Surprisingly, although TO mode showed larger variation in intensity 
and width compared to the LO mode (in fig.~2a,b), the overall shift 
is larger for LO mode ($\sim$8~cm$^{-1}$). With LO being a more surface 
sensitive mode, due to the Fr\"{o}hlich interactions \cite{nash}, 
lower intensity and FWHM of LO mode suggest that the lattice has 
undergone less modifications on the surface than in bulk. This, however, 
will lead to a larger stress on the surface. Higher shifts in LO mode 
show that this is indeed happening. Moreover, the shift being towards 
the lower wavenumbers suggests that the surface is under large tensile 
stress. Similar large shifts, in LO mode, towards lower wave numbers have
also been observed for InP Quantum Dots \cite{seong}. 

Figure~3 shows the  variations in intensity, width and position of L$^-$ 
plasmon-phonon coupled (LOPC) mode as a function of ion fluence. The 
intensity of L$^-$  mode increases upto the fluence of 
$1\times10^{13}~ions/cm^2$. For higher fluences, L$^-$ feature is not 
visible. As seen in Fig.~1, the c/a transition occurs in InP(111) lattice 
at the fluence $1\times10^{14}~ions/cm^2$. The disappearance of L$^-$ 
mode may reflect the capture of free carriers by defects due to the 
creation of amorphised zones after ion implantation. In Fig~3b, the 
FWHM of L$^-$ mode is seen to decrease with increasing fluence. This 
mode, thus, becomes sharper and more intense with increasing fluence. 
This is as expected since LOPC can become stronger with increasing 
carrier density.  The position of L$^-$ mode shifts towards lower 
wave numbers with the increasing fluence. As expected, the frequency
of L$^-$ mode asymptotically approaches TO frequency \cite{abs}.
With L$^-$ occurring at 320~cm$^{-1}$ for the virgin sample and
at 316~cm$^{-1}$ after the fluence of  $1\times10^{13}~ions/cm^2$,
the overall shift is about 4~cm$^{-1}$ for this mode. 

Figure~4 shows the second order Raman spectra of InP(111) after implantation 
with Sb ions at various fluences. For the virgin InP(111) a distinct triplet at 
617, 650 and and 682~cm$^{-1}$ is observed (fig.~4a). These features correspond 
to the 2TO, LO+TO and 2LO modes, respectively. The evolution of all these features, 
as a function of ion fluence, are displayed in fig.~4. It has been shown that 
the scattering volume for the second-order modes comprises the volume effectively 
probed by the exciting laser and hence these modes better reflect the nature of 
lattice modification than first order modes in zinc-blende (100) structures 
\cite{artus1}. Figure~5 shows the variations in the intensity, width and position 
of second order TO and LO modes, as a function of ion fluence. Figure~5a shows 
the intensity of 2TO and 2LO modes, at various fluences, normalized with respect 
to the respective intensities from the virgin sample. For both the modes, we 
observe a reasonable decrease in intensity at the fluence of 
$1\times10^{11}~ions/cm^2$. This result is in contrast to previous studies after 
keV implantation in InP(100) where no significant modifications in second order 
Raman modes were observed below the fluence of $5\times10^{12}~ions/cm^2$ 
\cite{ibanez,cusco}. For both 2LO and 2TO  modes, we observe a decrease in 
the intensity with increasing fluence. The reduction of intensity in Fig.~5a, 
like first order, 
is again larger for the 2TO mode than for 2LO. Although the overall decrease 
in intensity for LO and 2LO modes is similar, 2TO modes exhibit much less 
decrease compared to TO mode. At 
$1\times10^{13}~ions/cm^2$, we observe that for TO and 2TO modes the intensities are 
respectively about 25\% and 50\%  of the intensity from virgin sample. Figure~5b 
shows the widths (FWHM) of 2TO and 2LO features. The widths of 2LO and 2TO modes,
for the virgin InP(111), are observed to be 10 and 15~cm$^{-1}$, respectively.
These widths are slightly smaller than those observed for virgin InP(100) where
widths for LO and TO mode were found to be 12 and 16~cm$^{-1}$, respectively
\cite{artus2}. Although 2LO mode is slightly narrower than LO mode for the 
virgin sample 
as well as after implantation, the width of 2TO mode is larger than TO mode 
for the virgin sample. After implantation, the widths of TO and 2TO modes 
appear similar. Again like first order TO mode, 2TO mode shows larger and 
more rapid increase in width compared to 2LO mode.  The behavior of the shifts 
in position ($\omega$) of the second order modes is, however, different and 
unlike first order where LO mode exhibited larger shifts compared to TO mode, 
2LO mode shows very little change in its position. Figure~5c 
shows that 2LO mode experiences only a very slight shift ($\sim$2~cm$^{-1}$) towards 
the lower wave numbers after the fluence of $1\times10^{13}~ions/cm^2$ whereas 
2TO mode exhibits a large shift of about 7~cm$^{-1}$ (Fig.~5d). The shifts seen 
for 2TO mode are similar to those seen for TO mode in Fig.~2d. The modifications 
in the second-order Raman modes after implantation, like first-order, are associated 
with the degradation of the crystalline order in the InP(111) lattice. 

 The second order optic phonon spectra also contains a combination mode of 
LO+TO. Figure~6 shows the evolution of this combination mode as a 
function of ion fluence. The overall intensity of the mode initially increases
upto the fluence of $1\times10^{12}~ions/cm^2$ but decreases for higher fluences.
For $1\times10^{14}~ions/cm^2$, where c/a transition occurs, and higher fluences 
the combination LO+TO mode, is no more visible. The FWHM of the combination
mode is 15~cm$^{-1}$ for the virgin InP(111). This width is larger than the 
linewidth of combination mode (9~cm$^{-1}$) observed for virgin InP(100) 
\cite{artus2}. Surprisingly, though combination mode, seen here, is wider 
than the previous study \cite{artus2}, the 2LO and 2TO modes are narrower 
in the present study.
The linewidth of the combination mode increases with fluence becoming
30~cm$^{-1}$ after the fluence of $1\times10^{13}~ions/cm^2$ (Fig.~6b). 
The position ($\omega$) of this mode, however, does not show much shift with 
increasing ion fluence.

 During implantation, the lattice ions are displaced, creating 
defects and disordered regions in the process. With increasing 
lattice disorder, the phonon coherence length is reduced and
{\bf k}=0 selection rule is relaxed, giving rise to measurable
shifts and asymmetric broadening of the Raman peaks \cite{tiong}.
Due to translational symmetry breakdown, PCM developed by Richter 
et al. \cite{ley} can be used to evaluate the phonon confinement 
length or the average size of the undamaged crystalline regions.  
Assuming a constant correlation length {\it L} in the scattering volume, 
the intensity of the first order Raman band in the scattering volume 
is given by

$$I(\omega)=\int^{2\pi/a_0}_{0}\frac{\left|C(q)\right|^2 4\pi 
q^2dq}{[\omega-\omega(q)]^2+(\Gamma_0/2)^2}~~~~~~~~(1)$$

\noindent where $a_0$ is the lattice constant of InP (5.586~$\AA$).  
$\Gamma_0$ is the Raman intrinsic line width of the crystalline InP 
having values of 10.5 and 7.6 $cm^{-1}$ for TO and LO modes, 
respectively. The weight factor C(q) for the scattering with 
wave vector q is given by

$$\left|C(q)\right|^2=exp(\frac{-q^2L^2}{16\pi^2})~~~~~~~~(2)$$

\noindent and the phonon dispersion relation is given by \cite{sui}

$$\omega(q)=\omega_A-\omega_B(q/q_0)^2~~~~~~~~(3)$$

\noindent where $\omega_A$ is the wave vector of the first order Raman 
band in the absence of disorder effects and $q_0=\frac{2\pi}{a_0}$. 
Neutron scattering data provides $\omega_B$ to be 42 and 68~cm$^{-1}$, 
respectively, for LO and TO modes \cite{borcherds}. 
 
 By fitting the experimental LO and TO Raman modes with PCM we have obtained 
the phonon coherence length, L, of the as implanted samples. Figure~7 shows 
the results of fittings of PCM model to LO mode of Raman spectra, along with 
respective L values, at various fluences. Raman LO mode from the virgin InP 
sample is symmetric with an infinite coherence length. A fluence of 
$1\times10^{11} ~ions/cm^2$, causes an asymmetricity in the LO mode which can 
be due to the presence of  some disordered  or amorphous regions in the InP 
sample  that are large enough for phonon confinement.  By utilizing PCM, a 
phonon coherence length of 66~$\AA$ is obtained, indicating the presence of 
undamaged InP zones with  average crystallite size of this dimension. The 
shape of Raman LO mode after the fluence of $1\times10^{12} ~ions/cm^2$ shows 
a very slight increase in asymmetricity compared to that observed at 
$1\times10^{11} ~ions/cm^2$ and displays a coherence length of 56~$\AA$. 
A highly asymmetric Raman LO mode is observed after the fluence of 
$1\times10^{13} ~ions/cm^2$. The phonon coherence length at this stage is 
only 35~$\AA$. Thus, results after PCM fitting, for the fluences  
$1\times10^{11} ~ions/cm^2$ to $1\times10^{13} ~ions/cm^2$, suggest the 
presence of nanometer sized undamaged crystalline regions in the as-implanted 
InP matrix. Thus, a coexistence of nanocrystalline structures as well as
amorphous InP matrix takes place at these fluences. For higher fluences the 
InP lattice becomes completely amorphous. It is likely that the small amorphous 
zones inflate in size or some new amorphous zones get created with increasing
fluence that finally overlap leading to total amorphization of the lattice.  

In Fig.~7 we observe a coherence length of 35~$\AA$ at 
$1\times10^{13} ~ions/cm^2$. At this fluence, first order LO mode displays 
an overall shift of 8~cm$^{-1}$ and a width of 21~cm$^{-1}$ (see Fig.~2). 
Surprisingly, keV implantation at $1\times10^{13} ~ions/cm^2$ in InP(100), 
also leads to a similar coherence length ($\sim$36~$\AA$)\cite{soon}. However, 
shift, in LO mode, was only 1.3~cm$^{-1}$ whereas FWHM was 18~cm$^{-1}$.
For InP quantum dot (QD) of 35$\AA$ the shifts and FWHM of LO mode are found 
to be $\sim$4~cm$^{-1}$ and $\sim$20~cm$^{-1}$, respectively \cite{seong}. 
Although the width of LO mode in the present study as well as in studies 
by Yu etal.  \cite{soon} and Seong etal.\cite{seong} are similar, the shifts 
for same coherence length of 35 $\AA$, are very different.  Varying LO shifts 
may indicate varying order of associated stresses.  After MeV 
implantation of Sb in Si \cite{sdey2} the coherence length was found to be 
330~$\AA$ for a fluence of $1\times10^{13} ~ions/cm^2$.  Higher nuclear 
energy loss, S$_n$, for Sb in InP (1.9~keV/nm) compared to that in Si 
(1.2~keV/nm) will be responsible for the smaller nano-crystalline zones in 
InP at all fluences. 

The electron-phonon coupling strength can be estimated by measuring 
the normalized Raman intensity (I$_R$) of the 2LO phonon with respect 
to that of LO phonon. I$_R$ is considered approximately proportional
to the electron-LO phonon coupling \cite{ali,shi}. We have obtained I$_R$
for virgin InP(111) as well as after implantation at various fluences
and have shown them in Fig.~8. Furthermore, in Fig.~8 (top axis) we have also 
marked the coherence length, L, as obtained from the fitting results 
of PCM model (shown in Fig.~7) at all fluences. For virgin InP(111) 
we find I$_R$ of 0.20 in Fig.~8. The I$_R$ is seen to decrease as 
the fluence is increased. We observe  I$_R$ of 0.17, 0.16 and
0.13 for the fluences of $1\times10^{11}$, $1\times10^{12}$ and
$1\times10^{13} ~ions/cm^2$, respectively. From the PCM fitting 
results we had found that the coherence length of the crystalline
zones is 66$\AA$ at the fluence of $1\times10^{11} ~ions/cm^2$ 
(Fig.~7). It was also observed that as the fluence is increased, 
the coherence lengths decrease and become 56$\AA$ and 35$\AA$ 
for the fluences of $1\times10^{12}$ and $1\times10^{13}$, 
respectively. Thus, from Fig.~8 it is observed that the electron-
phonon coupling strength of nano-crystalline zones in InP decreases
as their sizes decrease. This behaviour is similar to the results seen
for InP Quantum dots \cite{seong}. However, the I$_R$ seen for InP 
QD (0.13 for 55$\AA$ and 0.08 for 35$\AA$) were very slightly smaller 
than those seen here for nano-crystalline zones created after 
MeV implantation. In the present study we observe that the electron-
phonon coupling, I$_R$, is largest for the virgin bulk-InP 
which also has biggest coherence length. With the decrease in 
coherence length, the electron-phonon coupling strength also 
decreases. The strength of coupling is about 1.5 times weaker in the
nano-crystalline zones, of 35$\AA$ size, than in the bulk -InP. 
This reduction in coupling is consistent with the increased overlap 
of electron and hole when there is substantial quantum confinement 
as also observed for bulk CdSe Quantum dots \cite{ali} where a 
20 times reduction in coupling strength, compared to bulk solid, 
was observed for a 45$\AA$ cluster. However, for InP quantum dots 
\cite{seong}, inexplicable reverse result with smallest I$_R$ 
of 0.06 for the bulk-InP was seen.

\section{Summary and Conclusion}

 InP(111) lattice, after MeV implantation, has been investigated by first
and second order Raman Spectroscopy. Evolution of both TO and LO modes 
as a function of ion fluence has been studied. The first order results are
influenced by the presence of a charge layer in the vicinity of surface. 
LO-phonon plasmon coupled mode, L$^-$, is seen to become stronger with 
increasing ion fluence. Results indicate coexistence of nanocrystalline zones
and amorphous regions in the implanted InP lattice. Electron- Phonon coupling
strength in nano-crystalline zones has been estimated and it reduces as the 
size of the crystalline regions decreases with increasing fluence. PCM has been 
applied to estimate the coherence length due to the confinement of phonons in the
implanted lattice and nano-crystalline zones of dimensions smaller than 66$\AA$ 
have been observed.

\section{Acknowledgments}

This work is partly supported by ONR grant no. N00014-97-1-0991. We would 
like to acknowledge the help of A. Pradhan in Raman scattering measurements.
We would  also like to thank Prof. S.N. Behera for useful discussions.

%\vspace*{1.5cm}
\newpage
\noindent{\bf \Large {Figures:}}
 
\vskip 0.3 in
\noindent Fig. 1: First order optical Raman spectra of virgin InP (a)
and 1.5 MeV Sb implanted InP at ion fluences of 
$1\times 10^{11} ~ions/cm^2$ (b), $1\times 10^{12} ~ions/cm^2$ (c), 
$1\times 10^{13} ~ions/cm^2$ (d), $1\times 10^{14} ~ions/cm^2$ (e), 
$5\times 10^{14} ~ions/cm^2$ (f) and $1\times 10^{15} ~ions/cm^2$ (g).   

\vskip 0.2 in
\noindent Fig. 2: Evolution of first order LO and TO modes:
The intensity of both LO and TO modes normalized with respect
to virgin (a), FWHM of both LO and TO modes (b) the shift in 
position of LO mode (c) and shift in position of TO mode (d) 
are shown as a function of ion fluence. Data for the virgin 
sample are also shown.

\vskip 0.2 in
\noindent Fig. 3: Evolution of L$^-$ mode: The intensity normalized
with respect to virgin (a), FWHM (b) and the shift in position (c) 
are shown as a function of ion fluence.  Data for the virgin sample 
are also shown.

\vskip 0.3 in
\noindent Fig.4: Second order optical Raman spectra of virgin InP (a)
and 1.5 MeV Sb implanted InP at ion fluences of 
$1\times 10^{11} ~ions/cm^2$ (b), $1\times 10^{12} ~ions/cm^2$ (c), 
$1\times 10^{13} ~ions/cm^2$ (d), $1\times 10^{14} ~ions/cm^2$ (e), 
$5\times 10^{14} ~ions/cm^2$ (f) and $1\times 10^{15} ~ions/cm^2$ (g)

\vskip 0.2 in
\noindent Fig. 5: Evolution of second order 2LO and 2TO modes:
The intensity of both 2LO and 2TO modes normalized with respect to
virgin (a), FWHM of both 2LO and 2TO modes (b) the shift in position 
of 2LO mode (c) and shift in position of 2TO mode (d) are shown as 
a function of ion fluence. Data for the virgin sample are also shown.

\vskip 0.2 in
\noindent Fig. 6: Evolution of LO+TO mode: The intensity normalized 
with respect to virgin (a), FWHM (b) and the shift in position (c) 
are shown as a function of ion fluence.  Data for the virgin sample 
are also shown.

 \vskip 0.2 in
\noindent Fig.7: Raman spectra of Sb implanted InP(111) at various
fluences fitted with PCM as described by Eq.(1). $\bullet$: experimental
data, --: phonon confinement model fit to data. L($\AA$) is the phonon
 coherence length as determined by the fit to the data. Fitting for the 
 virgin sample is also shown.

 \vskip 0.2 in
\noindent Fig.8: Normalized Raman intensity of second-order 2LO phonon 
with respect to that of first-order LO as a function of ion fluence. 
Value for the virgin bulk-InP is also shown. The coherence lengths (L), 
at all fluences, are also marked on the top axis. 
\newpage
\begin{figure}
\centerline{\epsfbox{fig1.eps}}
\large\bf\caption{}
\centerline {Paramanik et al.}
\end{figure}

\newpage

\begin{figure}
\centerline{\epsfbox{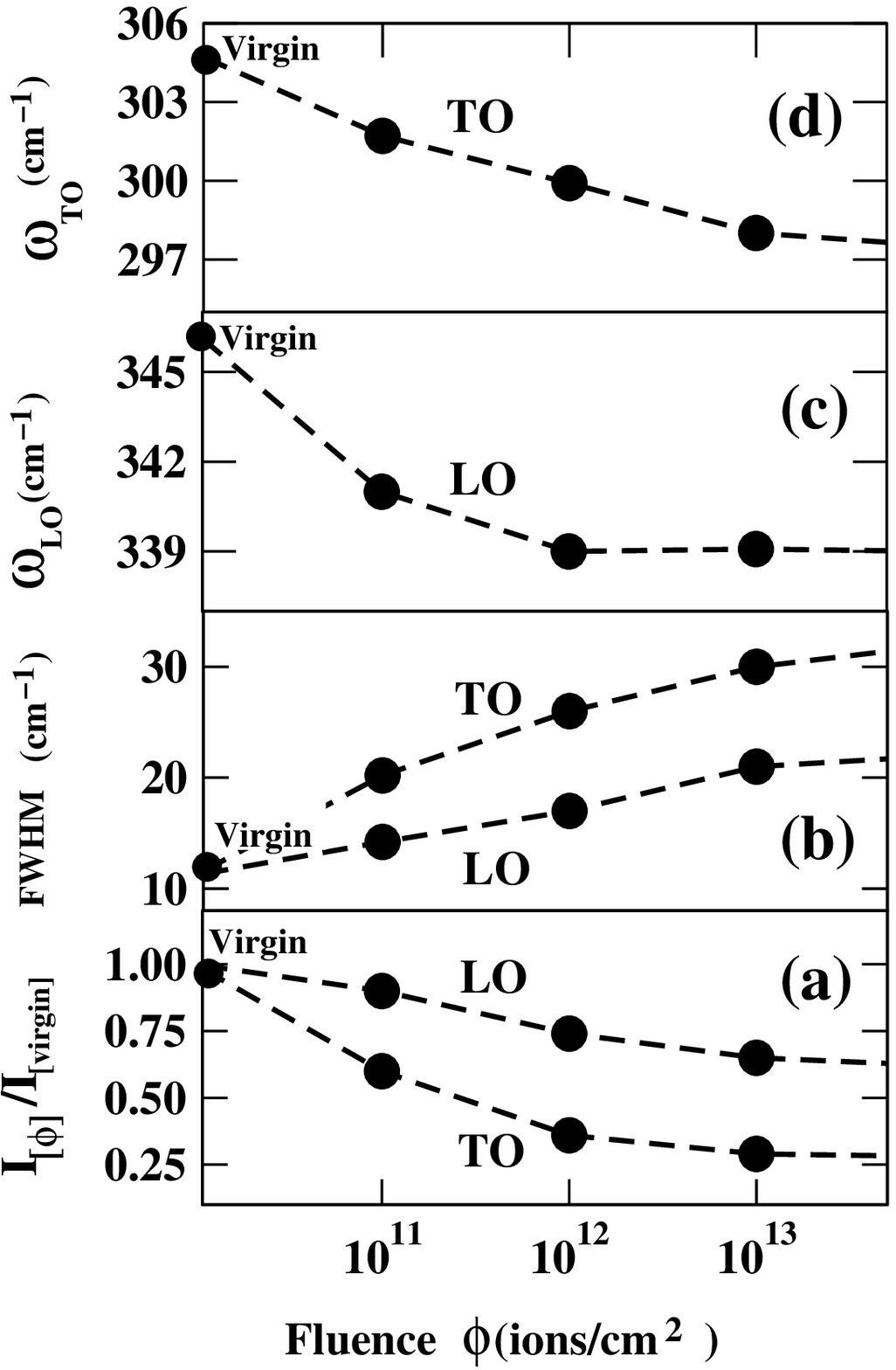}}
\large\bf\caption{}
\centerline {Paramanik et al.}
\end{figure}

\begin{figure}
\centerline{\epsfbox{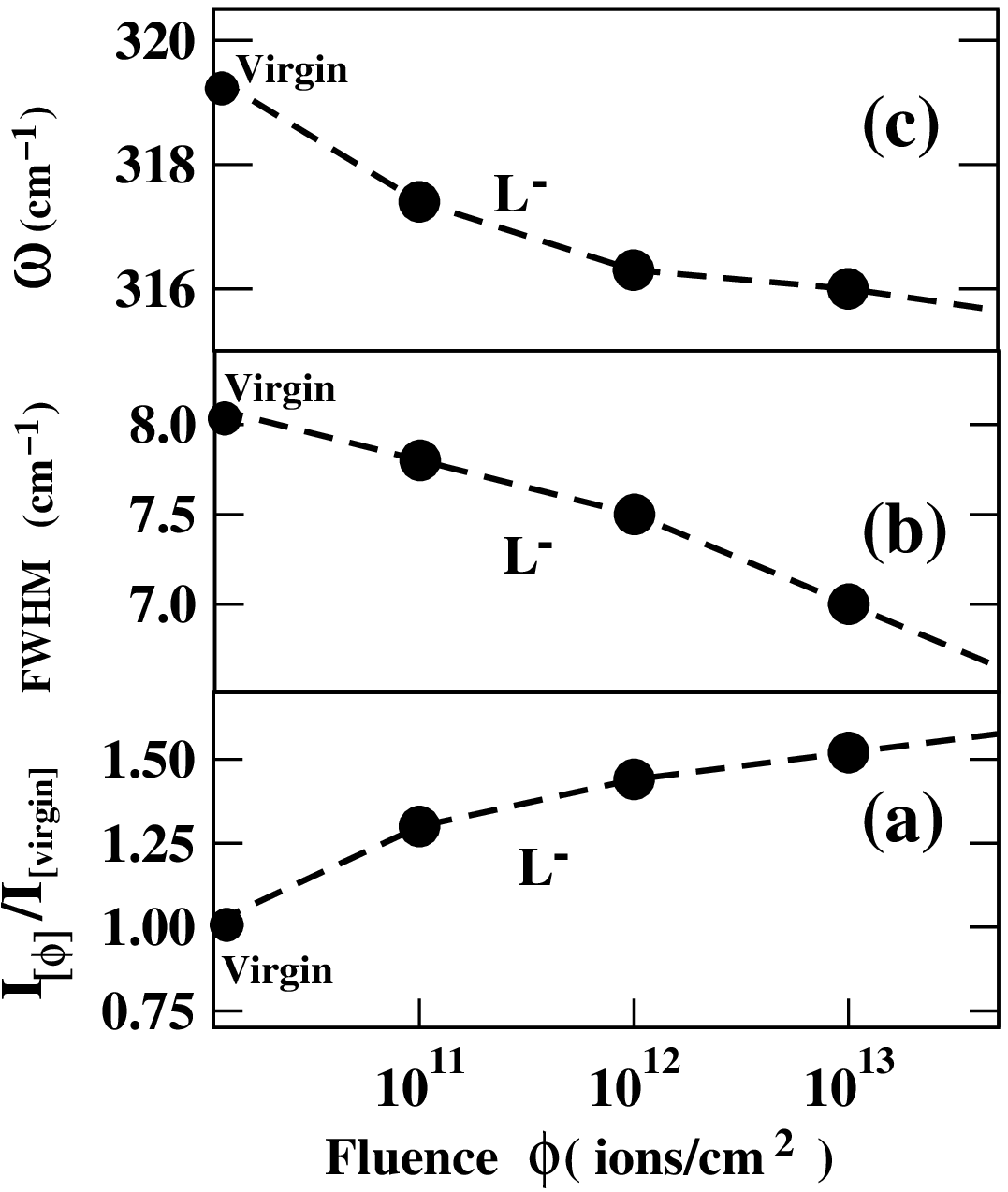}}
\large\bf\caption{}
\centerline {Paramanik et al.}
\end{figure}

\begin{figure}
\centerline{\epsfbox{fig4.eps}}
\large\bf\caption{}
\centerline {Paramanik et al.}
\end{figure}

\begin{figure}
\centerline{\epsfbox{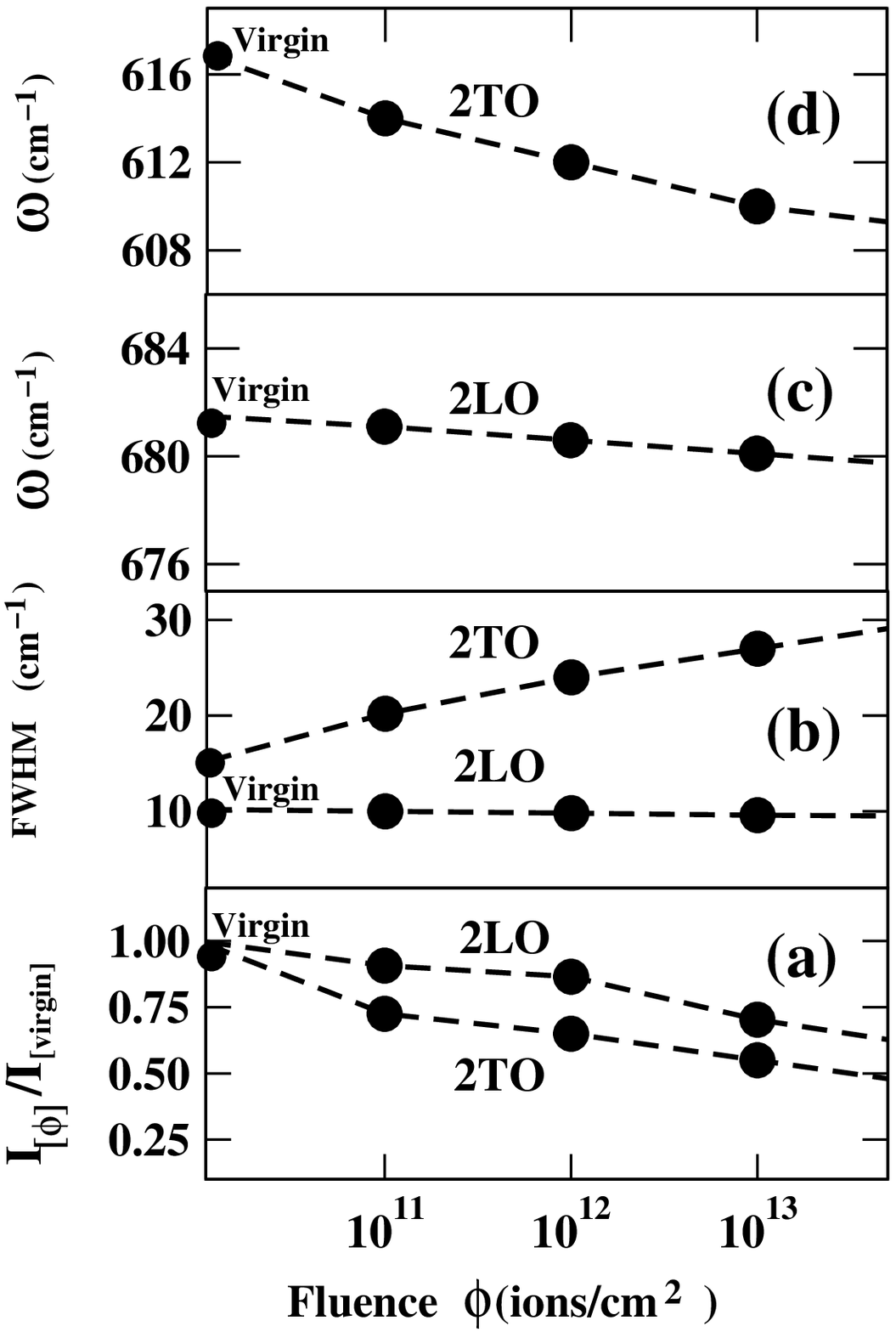}}
\large\bf\caption{}
\centerline {Paramanik et al.}
\end{figure}
\begin{figure}
\centerline{\epsfbox{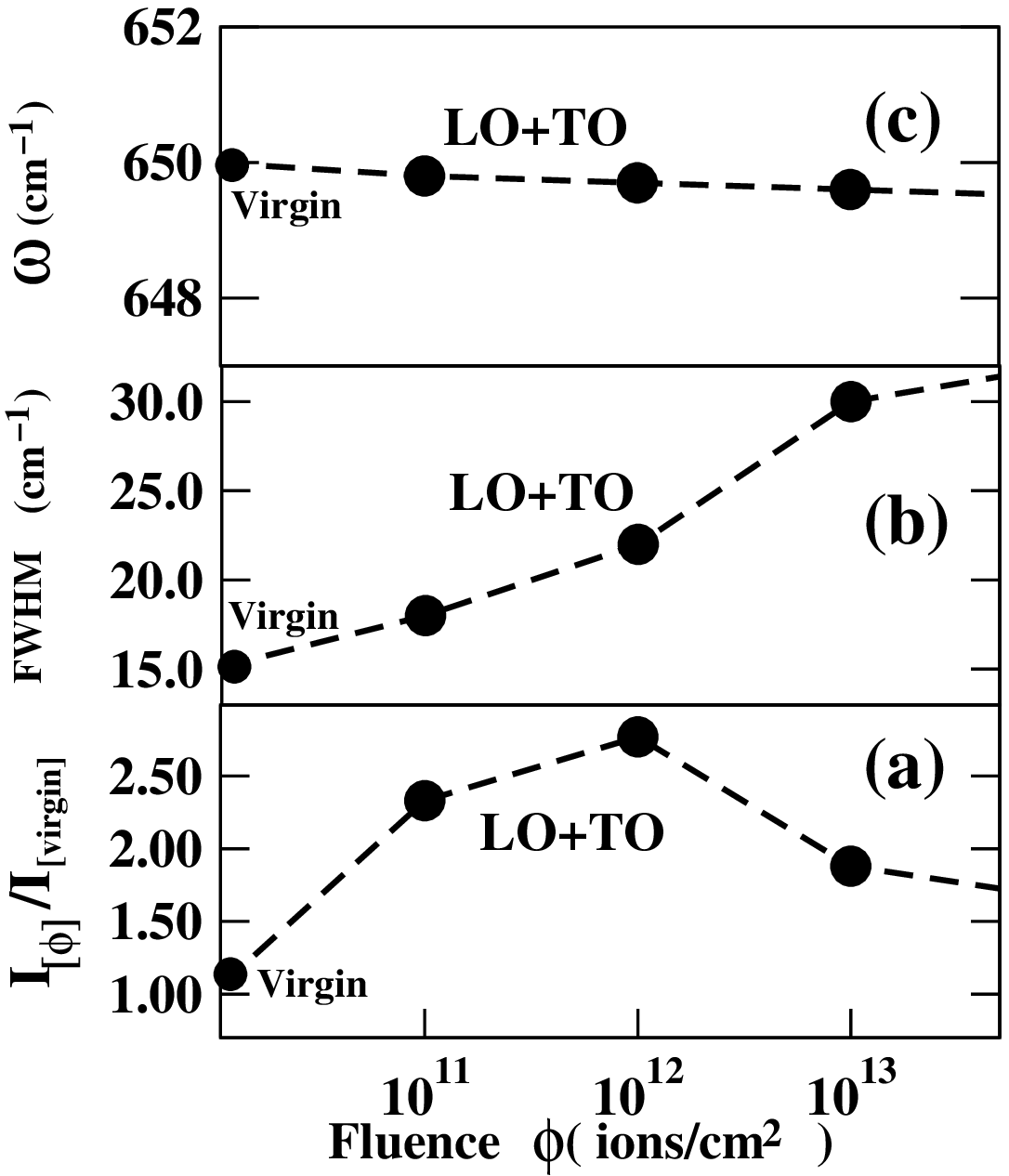}}
\large\bf\caption{}
\centerline {Paramanik et al.}
\end{figure}

\newpage

\begin{figure}
\centerline{\epsfbox{fig7.eps}}
\large\bf\caption{}
\centerline {Paramanik et al.}
\end{figure}

\begin{figure}
\centerline{\epsfbox{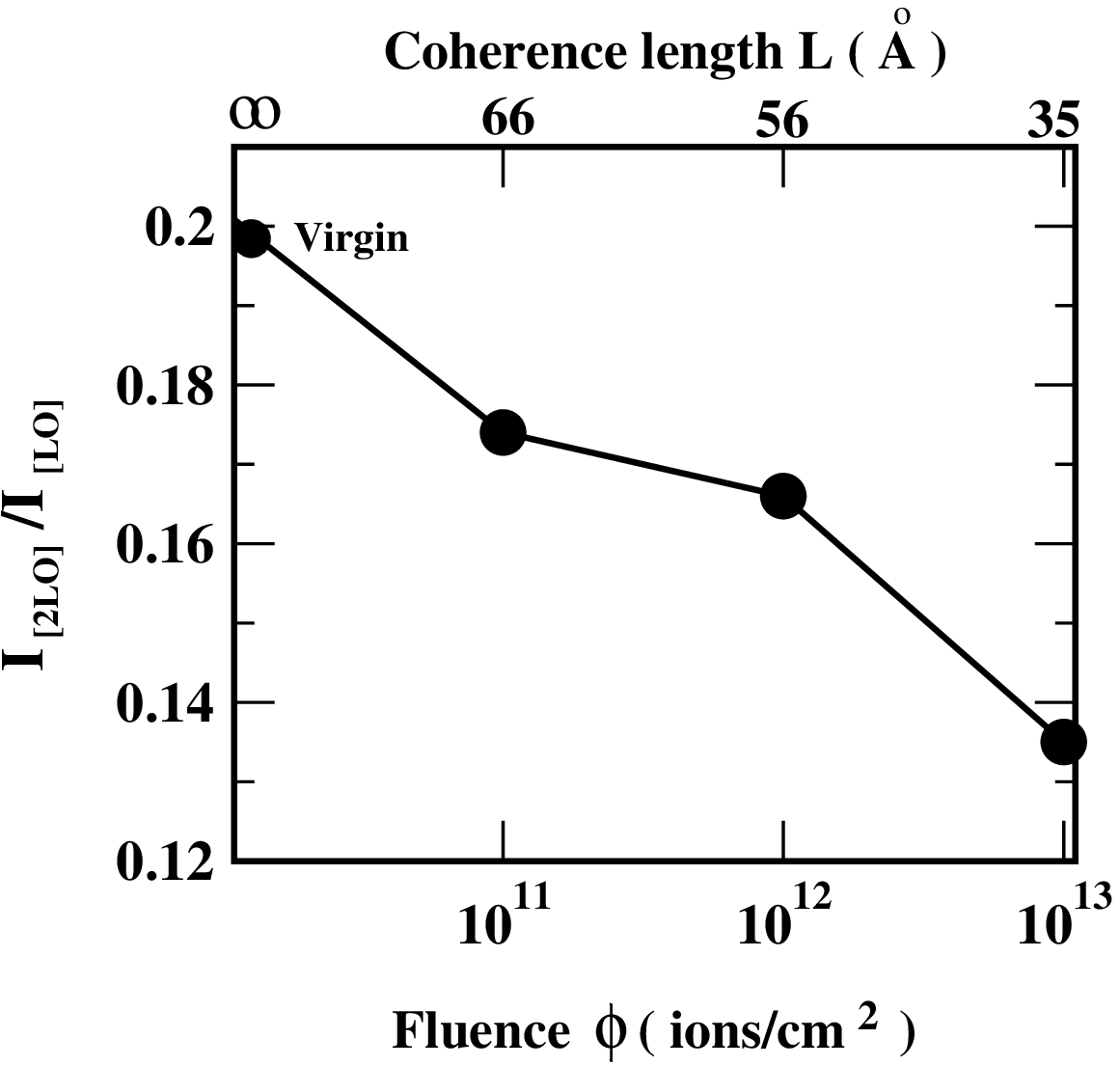}}
\large\bf\caption{}
\centerline {Paramanik et al.}
\end{figure}

\end{document}